\documentclass[twocolumn,showpacs,preprintnumbers,amsmath,amssymb]{revtex4}

\usepackage{graphicx}
\usepackage[dvips]{color}

\begin{document}


\title{A simple model of Feshbach molecules}

\author{Cheng Chin}

\affiliation{James Franck Institute and Department of Physics, University of Chicago, Chicago, IL 60637\\
 Institut f\"{u}r
Experimentalphysik, Universit\"{a}t Innsbruck, Technikerstr. 25,
6020 Innsbruck, Austria}

\date{\today}

\begin{abstract}
We present a two-channel model to describe the quantum state of two
atoms with finite-range interaction near a Feshbach resonance. This
model provides a simple picture to analytically derive the wave
function and the binding energy of the molecular bound state. The
results agree excellently with the measurements and multichannel
calculations. For small binding energies, the system enters a
threshold regime in which the Feshbach molecules are identical to
long range atom pairs in single channel. According to their
threshold behavior, we find Feshbach resonances can be classified
into two types.

\end{abstract}



\maketitle
\narrowtext\section{Introduction}

Formation and Bose-Einstein condensation (BEC) of molecules
\cite{mbec} have recently been achieved based on ultracold atoms
with magnetically-tuned Feshbach resonances \cite{feshbach}. In
these experiments, Feshbach coupling is induced by tuning a foreign
molecular state near the scattering continuum, which allows for an
efficient transfer of colliding atoms into molecules. This method
works for virtually all alkali atoms, and can create ultracold
molecules from various sources including Bose condensates
\cite{csmol}, degenerate Fermi gases \cite{k2bec}, or normal thermal
gases \cite{chi03, selim1}.

Feshbach molecules have special and unique properties. They
typically populate only one weakly-bound quantum state, and the
bound state can strongly couple to the scattering continuum via
Feshbach resonance. We may ask the following question: should
Feshbach molecules rather be considered as molecules in a specific
rovibrational state or as pairs of scattering atoms near the
continuum? This distinction is particularly crucial in the studies
of the BEC to BCS (Bardeen-Cooper-Schrieffer state) crossover in
degenerate Fermi gases, which call for a clarification of the
quantum nature of the Feshbach molecules \cite{strinati}.

Molecular states near Feshbach resonances have been recently
investigated based on sophisticated and complete two-body or
many-body theory \cite{stoof100, kokkelmans} and multi-channel
scattering calculations \cite{strinati, julienne28}. All works
suggest that the Feshbach molecule is generally a coherent mixture
of the foreign molecule in the closed channel and long-range atom
pair in the open scattering channel. Near resonances with large
resonance widths, the molecules can be well approximated as pairs in
the open channel. For narrow resonances, as suggested by numerical
calculation \cite{strinati, julienne28}, the closed channel
dominates and a short-range molecule picture is appropriate.

In this paper, we use a simple two-channel model to describe two
interacting atoms near a Feshbach resonance (Sec.~II). To account
for the finite interaction range of real atoms, we introduce a
spherical box potential, which allows us to analytically calculate
the molecular bound state in different regimes and their threshold
behavior (Sec.~III and Sec.~IV). Finally, we apply our model to
Feshbach molecules in recent Fermi gas experiments and to
characterize the associated Feshbach resonances (Sec.~V).

\section{Model}

We model the interaction of two identical, ultracold atoms with mass
$m$ based on an open channel $|o\rangle$ that supports the
scattering continuum and a closed channel $|c\rangle$ that supports
the foreign bound state. The wave function of the atoms is generally
expressed as $|\psi\rangle=\psi_o(r)|o\rangle+\psi_c(r)|c\rangle$,
where $\psi_o(r)$ and $\psi_c(r)$ are the amplitudes in the open and
closed channels, respectively, and $r$ is the inter-atomic
separation. We assume the interaction $(\hbar^2/m)\hat{v}$ is
described by a spherical box potential with an interaction range of
$r_0$, see Fig.~(1). For $r>r_0$, the potential energy of the open
channel is 0 and the closed channel $\infty$. For $r<r_0$, the open
(closed) channel has an attractive potential of $-\hbar^2q_o^2/m$
 $(-\hbar^2q_c^2/m)$, and a coupling term $\Omega=|\Omega|$ between the
channels. The wave function satisfies the Schr\"{o}dinger equation:

\begin{eqnarray}
E|\psi\rangle&=&\frac{\hbar^2}{m}(-\nabla^2+\hat{v})|\psi\rangle\\
\hat{v}&=&\left\{
\begin{array}{cc}-
\left(
\begin{array}{cc}
q_o^2 & \Omega \\
\Omega & q_c^2 \end{array} \right) & \mbox{ for $r<r_0$}  \\
& \\
\left(
\begin{array}{cc}
0 & 0 \\
0 & \infty \end{array} \right)& \mbox{ for $r>r_0$}. \end{array}
\right. \label{r0}
\end{eqnarray}

\begin{figure}
\includegraphics[width=2.2in]{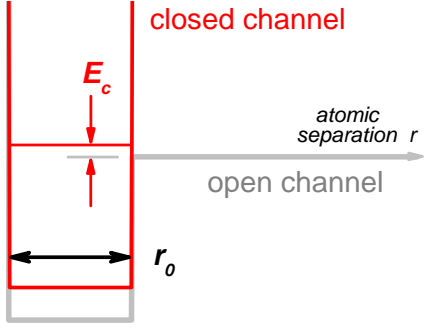}
\caption{A two-channel model for Feshbach resonances. We assume a
spherical box potential between two atoms with an interaction range
of $r_0$. A bound state with energy $E_c$ relative to the scattering
continuum is supported by the closed channel.} \label{fig1}
\end{figure}

%




The solution of the above equation for zero scattering energy $E=0$
can be expressed as:

\begin{eqnarray}
\mbox{for } r>r_0: |\psi\rangle&\propto&\frac{r-a}{r}|o\rangle, \\
\mbox{for } r<r_0: |\psi\rangle&\propto&\frac{\sin
q_+r}{r}|+\rangle+\frac{A\sin q_-r}{r}|-\rangle,
\label{r0}
\end{eqnarray}
where the scattering length $a$ and $A$ are constants, $q_{\pm}$ are
the ``eigen wave numbers" for $r<r_0$ associated with the eigen
states

\begin{eqnarray}
|+\rangle&=&\cos\theta |o\rangle+\sin\theta |c\rangle \nonumber \\
|-\rangle&=&-\sin\theta |o\rangle+\cos\theta |c\rangle, \label{r0}
\end{eqnarray}
and $\tan2\theta=2\Omega/(q_o^2-q_c^2)$.

Based on the boundary conditions $\psi_c(r_0)=0$ and
$\psi_o(r_0)/\psi_o'(r_0)=r_0(r_0/a-1)$, we get

\begin{eqnarray}
A&=&-\tan\theta\,\frac{\sin(q_+r_0)}{\sin(q_-r_0)} \\
\frac{1}{r_0-a}&=&\frac{q_+\cos^2\theta}{\tan{q_+}r_0}+\frac{q_-\sin^2\theta}{\tan{q_-}r_0}.
\label{r0}
\end{eqnarray}

\noindent The latter equation shows how in general, each channel
contributes to the scattering length.

 In cold atom systems, Feshbach resonances are, in most cases, induced
by hyperfine interactions or spin-spin interactions. Both
interactions are many orders of magnitude weaker than the relevant
short range exchange potential. It is an excellent approximation to
assume $\Omega\ll q_o^2, q_c^2$ and $|q_o^2-q_c^2|$. Hence, we have
$\theta\ll1$, $q_+\approx q_o$ and $q_-\approx q_c$.

In this limit, the closed channel contribution is significant only
when the foreign state is close to the continuum, in which case the
last term in Eq.~(7) diverges. Given the energy of the closed
channel state as $E_c=(\hbar^2/m)\epsilon_c$ and $\epsilon_c\ll
q_c/r_0$, the boundary condition
$\psi_o(r_0)=\sin\sqrt{q_c^2+\epsilon_c}\,r_0=0$ allows us to expand
the last term in Eq.~(7) as $-\gamma/\epsilon_c$. Here
$\gamma=2q_c^2\theta^2/r_0$ characterizes the Feshbach coupling
strength. To the same order of expansion, the middle term in Eq.~(7)
is a constant across the resonance and can be identified as
$(r_0-a_{bg})^{-1}$, where $a_{bg}$ is the background scattering
length. Equation (7) reduces to

\begin{eqnarray}
\frac{1}{r_0-a}&=&\frac{1}{r_0-a_{bg}}-\frac{\gamma}{\epsilon_c}.
\end{eqnarray}

\maketitle\narrowtext\section{Scattering length and the molecular
eigen state}




Experimentally, the relative energy between the continuum and the
bare state can be adjusted linearly by a magnetic field $B$-induced
Zeeman shift $-\mu B$, where $\mu=\mu_o-\mu_c$ and $\mu_o(\mu_c)$ is
the magnetic moment of the open(closed) channel. Replacing $E_c$ by
$E_c+\mu B$, we can rewrite Eq.~(8) in terms of the magnetic field
as


\begin{equation}
\frac{a-r_0}{a_{bg}-r_0}=1+\frac{\Delta B}{B-B_{res}},\label{r0}
\end{equation}
where the resonance width $\Delta B$ and the resonance position
$B_{res}$ are given by

\begin{eqnarray}
\Delta B&=&-\mu^{-1}(\frac{\hbar^2}{m})\gamma(a_{bg}-r_0) \\
B_{res}&=&-\mu^{-1}E_c+\Delta B. \label{r0}
\end{eqnarray}

Several interesting features are shown here. First of all, we find
the resonance width is proportional to both the Feshbach coupling
$\gamma$ and the background scattering properties $a_{bg}-r_0$. The
latter dependence is due to the fact that the scattering amplitude
at short range is proportional to the scattering length. A larger
short range scattering amplitude leads to a stronger coupling to the
closed channel.

Secondly and importantly, the resonance position is offset by
exactly $\Delta B$ relative to the crossing of the bare state and
the continuum, $B=-\mu^{-1} E_c$, see Eq.~(11). For a positive
scattering length $a_{bg}>r_0$, this shift is negative $\Delta B<0$.
This feature leads to the ``renormalization" of the Feshbach
resonance location discussed in Ref.~\cite{kokkelmans}.

To understand the origin of the resonance shift, we should return to
Eq.~(8). The divergence of the scattering length occurs when the
open channel contribution (middle term) is exactly canceled by the
closed channel one (last term). For systems with large background
scattering lengths $|a_{bg}|\gg r_0$ and strong Feshbach couplings
$\gamma$, this cancelation can occur even when the bare state is far
away from the continuum. A large resonance shift then results.

Now we turn to the binding energy of the molecules. Assuming a bound
eigen state $|\psi_m\rangle$ exists near the continuum at
$E=-E_m=-(\hbar^2/m)\epsilon_m$, where $E_m>0$ is the binding
energy, we can determine $E_m$ by following essentially the same
calculation as Eq.~(1)-(7). The equivalence of Eq.~(7) gives

\begin{eqnarray}
-\sqrt{\epsilon_m}=\frac{\bar{q}_+\cos\theta^2}{\tan{\bar{q}_+}r_0}+\frac{\bar{q}_-\sin\theta^2}{\tan{\bar{q}_-}r_0},
\label{r0}
\end{eqnarray}
where $\bar{q}_\pm=(q_\pm^2-\epsilon_m)^{1/2}$.

Assuming $\theta\ll1$ and the bound states in both channels are
close to the continuum, namely, $|a_{bg}|\gg r_0$ and $|\epsilon_c|
\ll q_o/r_0$, we can expand the two terms on the right side of
Eq.~(12) to leading order as $(r_0-a_{bg})^{-1}$ and
$-\gamma(\epsilon_c+\epsilon_m)^{-1}$, respectively. Equation~(12)
then reduces to

%


\begin{equation}
(\epsilon_m+\epsilon_c)(\sqrt{\epsilon_m}-\frac{1}{a_{bg}-r_0})=\gamma.
\label{r0}
\end{equation}

This result shows the evolution of the eigen state near the
resonance. Similar result is obtained in Ref.~\cite{kokkelmans}
based on a contact potential. We can immediately see that in the
absence of the Feshbach coupling $\gamma=0$, the solutions of
Eq.~(13) are $\epsilon_m=-\epsilon_c$ and
$\epsilon_m=(a_{bg}-r_0)^{-2}$ (for $a_{bg}>r_0$), which exactly
correspond to the bare bound states in the closed channel and the
open channel (for $a_{bg}>r_0$), respectively.

In the presence of the Feshbach coupling $\gamma>0$, Eq.~(13)
suggests an ``avoided level crossing"-like energy structure, see
Fig.~(2), which also illustrates the resonance position shifts. The
level crossing, however, is not hyperbolic as it is in a two-level
system. In particular, at small binding energies, the bound state
energy approaches the continuum quadratically, see Fig.~(2) inset.
Far below the continuum, the bound state approaches the bare state
in the closed channel.

\begin{figure}
\includegraphics[width=2.8in]{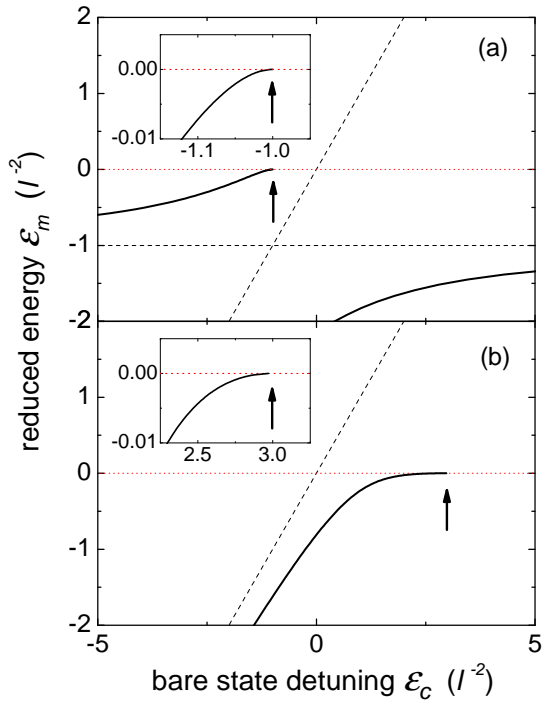}
\caption{Generic energy structure of Feshbach molecules. Based on
Eq.~(13), we show the reduced energies of the continuum (dotted
lines), bare molecular states (dashed lines), and the dressed
molecular state ($\epsilon_m$, solid lines). We assume (a)
$a_{bg}-r_0=l$ and $\gamma=l^{-3}$ and (b) $a_{bg}-r_0=-3l$ and
$\gamma=l^{-3}$, where $l\gg r_0$ can be any relevant length scale.
Arrows mark the offset resonance positions. Insets show the
threshold behavior of the bound state.} \label{fig2}
\end{figure}

To better quantify the role of the open and closed channel, we can
write the wave function of the eigen state as

\begin{eqnarray}
r>r_0: |\psi_m\rangle&\propto& \frac{e^{-\sqrt{\epsilon_m}r}}{r} |o\rangle \\
r<r_0: |\psi_m\rangle&\propto& \frac{\sin
\bar{q}_+r}{r}|+\rangle+\frac{A_m\sin \bar{q}_-r}{r}|-\rangle,
\label{r0}
\end{eqnarray}
and, with our approximations, $A_m$ satisfies

\begin{eqnarray}
A_m^2=\frac{2\gamma}{r_0}(\epsilon_c+\epsilon_m)^{-2}=\frac{\tan^2\phi}{\sqrt{\epsilon_m}r_0},
\label{r0}
\end{eqnarray}
and the mixing angle $\phi$ is defined below.

We show in Eq.~(15) that the eigen state generally occupies both the
closed channel and open channel. We can introduce a mixing amplitude
$\sin\phi$ as the amplitude in the closed channel

\begin{equation}
|\psi_m\rangle=\sin\phi |c\rangle+\cos\phi |o\rangle. \\
\label{r0}
\end{equation}

The mixing fraction $\sin^2\phi$ can be evaluated by a direct
integration of the closed channel wave function. Alternatively,
noticing that the mixing also leads to a dependence of the eigen
state on the bare state, we can also derive $\sin^2\phi$ from the
dependence of $E_m$ on $E_c$, or from the magnetic moment of the
Feshbach molecule $\mu_m=\partial E_m/\partial B$. All methods lead
to the same result

\begin{eqnarray}
\sin^2\phi&=&\int|\psi_c(r)|^2d^3r=-\frac{\partial E_m}{\partial E_c}=\frac{\mu_o-\mu_m}{\mu}\\
          &=&\frac{2\sqrt{\epsilon_m}\gamma}{(\epsilon_c+\epsilon_m)^2+2\sqrt{\epsilon_m}\gamma}.
\label{r0}
\end{eqnarray}

\maketitle\narrowtext\section{Threshold Regime in open channel}

%

Despite the seemingly complex equations shown in previous sections,
the Feshbach molecules are simple and universal near the scattering
continuum. Expanding Eq.~(13) with small $E_m$ and using
Eq.~(9)-(11), we find the binding energy of the Feshbach molecules
has a simple dependence on the scattering length and increases
quadratically in magnetic field near the resonance, namely,

\begin{eqnarray}
E_m&=&\frac{\hbar^2}{m(a-r_0)^2} \\
&=&\frac{\mu^2(B-B_{res})^2}{4E^*}, \label{r0}
\end{eqnarray}

\noindent where $E^*=(\hbar/4m)\gamma^2(a_{bg}-r_0)^4$.




Equation (20) shows identical dependence on scattering length and
interaction range as of single channel molecules in the threshold
regime \cite{flambaum}. Furthermore, taking the limit
$\epsilon_m\rightarrow 0$ in Eq.~(14) and (19), we find the
molecular wave function here is purely in the open channel. Its
spatial extent is determined by quantum uncertainty, $\langle r
\rangle \sqrt{m E_m}=\hbar/2$, and can be much larger than the
interaction range $\langle r\rangle=a/2\gg r_0$. In this limit, the
Feshbach molecules are identical to long-range atom pairs in a
single open channel.

By expanding Eq.~(19) at small $\epsilon_m$ and using Eq.~(13), we
find the closed channel fraction can be expressed as
\begin{equation}
\sin^2\phi=\sqrt{\frac{E_m}{E^*}}-(1+\delta)\frac{E_m}{E^*}+...,
\label{r0}
\end{equation}

\noindent where $\delta=\gamma(a_{bg}-r_0)^3$.

From Eq.~(22), we see that $E^*$ provides the leading order
estimation of the closed channel admixture. When $E_m\ll E^*$ or
$E_m\ll E^*/|\delta|$ (this condition applies when $\delta\ll-1$),
the Feshbach molecule is purely in the open channel. As expected,
the threshold regime is wider for resonances with larger $\gamma$
and $|a_{bg}-r_0|$.

We can further determine the ``open channel-dominated" regime by
setting $\sin^2\phi<\frac12$ in Eq.~(19). For resonances with small
$|\delta|\ll 1$, this condition corresponds to $E_m<E^*$, which, in
terms of magnetic field, maps to only a small fraction of
$|\delta|\Delta B$ near the resonance $B_{res}$. For resonances with
large $|\delta|\gg1$, the open channel dominates when
$E_m<(\gamma/2)^{2/3}$, which covers the full resonance width when
$a_{bg}-r_0<0$, and covers the entire upper branch of the bound
state when $a_{bg}-r_0>0$.

Based on the range of the single channel regime, we suggest the
broad(narrow) resonances be defined as those with $|\delta|\gg1$
($|\delta|\ll1$). Within the width of the Feshbach resonance, the
molecules associated with a broad (narrow) resonance are better
described as long range pairs in the open channel (short range
molecules in the closed channel). We note that this definition is
purely based on two-body physics.

\maketitle\narrowtext\section{Feshbach molecules in $^6$Li and
$^{40}$K}

\begin{table}
\caption{Parameters of the $^6$Li and $^{40}$K Feshbach resonances.
Interaction range $r_0$ is derived from Ref.~\cite{flambaum}, see
text. Feshbach coupling $\gamma$ is derived from Eq.~(10). $a_0$ is
Bohr radius and $\mu_B$ is Bohr magneton.}

\begin{tabular}{cccccccc}\hline
   & $r_0 (a_0)$ & $B_{res}$(G) & $\Delta B$(G) & $a_{bg}(a_0)$ & $\mu(\mu_B)$ & $\gamma^{-1/3} (a_0)$ & Ref.\\ \hline
  $^6$Li & 29.9 & 834.15  & 300 & -1405 & 2.0 & 101 & \cite{boundbound}  \\
  $^{40}$K & 62 & 224.21  & -9.77 &  174 & 2.5 & 67 & \cite{kmeasure}  \\
  $^6$Li & 29.9 & 543.26 & -0.1 & 61.6 & 2.1 & 400 & \cite{hulet} \\ \hline
\end{tabular}
\label{Table1}
\end{table}

\begin{figure}
\includegraphics[width=2.8in]{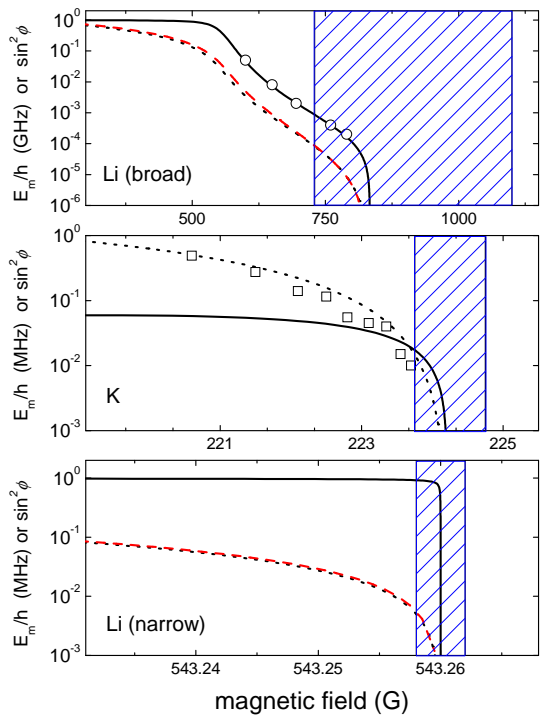}
\caption{Binding energies  $E_{m}$ (dotted lines) and mixing
fractions $\sin^2\phi$ (solid lines) of the molecules near the
$^6$Li and $^{40}$K Feshbach resonances. The curves are calculated
from Eq.~(13), Eq.~(19) and the parameters in Table 1. Binding
energies from multi-channel calculation \cite{paul} (dashed lines),
from JILA group measurement \cite{kmeasure} (open square) and the
mixing fractions measurement from Rice group \cite{hulet2005} (open
circles) are shown for comparison. The shaded areas indicate the
typical BEC-BCS crossover regimes, $|a|>3000 a_0$.} \label{fig2}
\end{figure}

Finally, we apply our model to the $^6$Li$_2$ and $^{40}$K$_2$
Feshbach molecules created in recent BEC-BCS experiments
\cite{mbec}. These molecules are stable near the resonance and both
the molecular binding energies and the scattering lengths have been
well measured and studied \cite{boundbound, kmeasure}.

To model the interaction of atoms, we adopt $B_{res}$, $\Delta B$,
$B_{bg}$ and $\mu$ from recent measurements and numerical
calculations. To account for the finite range of the atomic
interaction, which at low temperatures is determined by the van der
Waals potential of $\sim-r^{-6}$, we choose the interaction range
$r_0$ in our model to be the mean scattering length $\bar{a}$
defined in Ref.~\cite{flambaum}. This choice ensures the same
behavior of the scattering length in the threshold regime
\cite{flambaum}. All parameters are given in Table~I.

In Fig.~(3), we show the calculated binding energy $E_m$ and the
mixing fraction $\sin^2\phi$ of the Feshbach molecules for the two
$^6$Li resonances and one $^{40}$K resonance. The results agree very
well with the multi-channel calculation \cite{paul} and the
measurements on molecular binding energy \cite{kmeasure,
boundbound}, magnetic moment \cite{selim1} and mixing fraction
\cite{hulet2005}.

Both the Li resonance at 834G and the K resonance are broad with
$\delta=-2900$ and $4.7$, respectively. The open channel-dominated
regimes of $E_m<E^*/|\delta|=h$~210~MHz for the 834~G Li resonance
is also larger than the Fermi energy of $E_F\approx h$~20~kHz in the
experiments.(Here the Fermi wave number is
$k_{F}\sim(3000a_0)^{-1}$.) For the K resonance, the full upper
branch of the molecular state is open channel dominated with mixing
fractions less than $10\%$. Therefore, we conclude the open channel
description of these Feshbach molecules in the crossover regime to
be a good approximation.

For the narrower Li resonance at $\sim$543~G, we obtain
$\delta\sim0.0005$ and $E^*=h$~31~Hz $\ll E_{F}=h$~20~kHz. This
indicates an extremely narrow open channel regime of less than
50~$\mu$G near the resonance, where the gas parameter is still over
$k_{F}a=20$. Crossover experiments based on these Feshbach molecules
cannot be described by open channel atom pairs and may lead to
qualitatively different physics. We attribute the large difference
between the two Li resonances to their different couplings $\gamma$
and very different background scattering length $a_{bg}$, see Table
I.

In the above discussions, we note that Fermi energy $E_F$ is an
external parameter which depends on the density of the sample.
Whether the molecules in the crossover regime can be described by
single channel strongly depends on the density. The $\delta$
parameter, however, provides a better and independent measure to
classify Feshbach resonances. We find that the two Feshbach
resonances in $^6$Li are the two extremes of broad and narrow
resonances with $\delta=-2900$ and $\delta=0.0005$.

In summary, the two-channel model provides a simple picture to
understand the molecular state near the Feshbach resonances. The
analytic results of the molecular binding energy and mixing fraction
on $^6$Li and $^{40}$K agree with the measurements and other
sophisticated calculations very well. Based on the threshold
behavior of the bound state, we suggest a dimensionless parameter to
assess the ``broadness" of the Feshbach resonance.

\section*{Acknowledgements}
We thank P.S. Julienne and N. Nygaard for stimulating discussions
and R. Grimm's lithium and cesium groups in Innsbruck for the
support during our visit. The author is partially supported by the
Lise-Meitner program of the Austrian Science Fund (FWF).


\begin{references}
\bibitem{mbec}
S. Jochim, M. Bartenstein, A. Altmeyer, G. Hendl, S. Riedl, C. Chin,
J. Hecker Denschlag and R. Grimm, Science {\bf 302}, 2101 (2003); M.
Greiner, C.A. Regal, D.S. Jin, Nature {\bf 426}, 537 (2003); M.
Zwierlein, C.A. Stan, C.H. Schunck, S.M.F. Raupach, S. Gupta, Z.
Hadzibabic, and W. Ketterle, Phys. Rev. Lett. {\bf 91}, 250401
(2003).

\bibitem{feshbach}
E. Tiesinga, B.J. Verhaar, and H.T.C. Stoof, Phys. Rev. A {\bf 47},
4114 (1993); S. Inouye, M. Andrews, J. Stenger, H.-J. Miesner, S.
Stamper-Kurn, and W. Ketterle, Nature {\bf 392}, 151 (1998).

\bibitem{csmol}
J. Herbig, T. Kraemer, M. Mark, T. Weber, C. Chin, H.-C. N\"{a}gerl,
and R. Grimm, Science {\bf 301}, 1510 (2003); S. D\"{u}rr, T. Volz,
A. Marte, and G. Rempe Phys. Rev. Lett. {\bf 92}, 020406 (2004);  K.
Xu, T. Mukaiyama, J.R. Abo-Shaeer, J.K. Chin, D. Miller, and W.
Ketterle, Phys. Rev. Lett. {\bf 91}, 210402 (2003).

\bibitem{k2bec}
M. Greiner, C. A. Regal, D. S. Jin, Nature {\bf 426}, 537 (2003).

\bibitem{chi03}
C. Chin, V. Vuleti\'{c}, A.J. Kerman, and S. Chu, Phys. Rev. Lett.
{\bf 90}, 033201 (2003).

\bibitem{selim1}
S. Jochim, M. Bartenstein, A. Altmeyer, G. Hendl, C. Chin, J. Hecker
Denschlag, and R. Grimm, Phys. Rev. Lett. {\bf 91}, 240402 (2003).

\bibitem{strinati}
S. Simonucci, P. Pieri and G.C. Strinati, Europhys. Lett. {\bf 69}
713 (2005).

\bibitem{stoof100} R.A. Duine and H.T.C. Stoof, Phys.
Rep. {\bf 396}, 115 (2004).

\bibitem{kokkelmans}
B. Marcelis, E.G.M. van Kempen, B.J. Verhaar, and S.J.J.M.F.
Kokkelmans, Phys. Rev. A {\bf 70}, 012701 (2004); S.J.J.M.F.
Kokkelmans, J.N. Milstein, M.L. Chiofalo, R. Walser, and M.J.
Holland, Phys. Rev. A {\bf 65}, 053617 (2002).

\bibitem{julienne28}
K. Goral, T. Koehler, S.A. Gardiner, E. Tiesinga, P.S. Julienne, J.
Phys. B {\bf 37}, 3457 (2004).

\bibitem{boundbound}
M. Bartenstein, A. Altmeyer, S. Riedl, R. Geursen, S. Jochim, C.
Chin, J. Hecker Denschlag, R. Grimm, A. Simoni, E. Tiesinga, C.J.
Williams, and P.S. Julienne,  Phys. Rev. Lett. {\bf 94}, 103201
(2005).


\bibitem{flambaum}
G.F. Gribakin and V.V. Flambaum, Phys. Rev. A {\bf 48}, 546 (1993).

\bibitem{paul}
A. Simoni, V. Venturi and P. S. Julienne, private communication.

\bibitem{kmeasure}
C.A. Regal and D.S. Jin, Phys. Rev. Lett. {\bf 90}, 230404 (2003);
C.A. Regal, C. Ticknor, J.L. Bohn, and D.S. Jin, Nature {\bf 424},
47 (2003).

\bibitem{hulet}
K.E. Strecker, G.B. Partridge and R.G. Hulet, Phys. Rev. Lett. {\bf
91}, 080406 (2003).

\bibitem{hulet2005}
G.B. Partridge, K.E. Strecker, R.I. Kamar, M.W. Jack, R.G. Hulet,
cond-mat/0505353.

\end{references}
\end{document}